\documentclass[twocolumn,prl,preprintnumbers,superscriptaddress]{revtex4-2}

\usepackage{graphicx, dcolumn, bm, amsmath, dsfont, xcolor, comment}

\makeatletter
\def\maketitle{
\@author@finish
\title@column\titleblock@produce
\suppressfloats[t]}
\makeatother

\begin{document}

\title{Noncooperative Quantum Networks}

\author{Yanxuan Shao}
\affiliation{Department of Physics and Astronomy, Northwestern University, Evanston, IL 60208}
\affiliation{Center for Network Dynamics, Northwestern University, Evanston, IL 60208}

\author{Jannik L. Wyss}
\affiliation{Department of Physics and Astronomy, Northwestern University, Evanston, IL 60208}
\affiliation{Center for Network Dynamics, Northwestern University, Evanston, IL 60208}
\affiliation{Department of Physics, University of Basel, Klingelbergstrasse 82, CH-4056 Basel, Switzerland}

\author{Don Towsley}
\affiliation{College of Computer Science, University of Massachusetts Amherst, Amherst, MA 01003}

\author{Adilson E. Motter}
\affiliation{Department of Physics and Astronomy, Northwestern University, Evanston, IL 60208}
\affiliation{Center for Network Dynamics, Northwestern University, Evanston, IL 60208}
\affiliation{Department of Engineering Sciences and Applied Mathematics, Northwestern University, Evanston, IL 60208 }
\affiliation{Northwestern Institute on Complex Systems, Northwestern University, Evanston, IL 60208}
\affiliation{Institute for Quantum Information Research and Engineering, Northwestern University, Evanston, IL 60208}

\begin{abstract}
Existing protocols for quantum communication networks usually assume an initial allocation of quantum entanglement resources, which are then manipulated through local operations and classical communication (LOCC) to establish high-fidelity entanglement between distant parties. It is generally held that the resulting fidelity would increase monotonically with the entanglement budget. Here, we show that for noncooperative LOCC protocols, the resulting fidelity may decrease as more entanglement is added to a network with non-pure states. This effect results from a quantum analog of selfish routing and constitutes a potential obstacle to the optimal use of resources in large quantum networks. 

\end{abstract}

\maketitle

\textit{Introduction} -- Quantum phenomena have been increasingly studied in network settings \cite{biamonte2019complex, nokkala2024complex} owing to a variety of applications, including quantum cryptography \cite{pirandola2020advances, bennett1984quantum}, 
distributed quantum computing \cite{main2025distributed, wei2025universal}, 
high-precision telescope systems \cite{khabiboulline2019optical, gottesman2012longer}, and the possibility of a global quantum internet \cite{wehner2018quantum, cacciapuoti2020quantum}.
These applications require entanglement between remote locations, which can be implemented via optical fibers using entangled photons. 
However, the photon transmission rate decays exponentially with distance \cite{simon2017towards}, necessitating
quantum-repeater-based entanglement manipulation to enable long-distance quantum communication \cite{sangouard2011quantum, briegel1998quantum, azuma2023quantum}. This is achieved
by converting multiple consecutive entangled pairs into a single direct end-to-end 
entanglement, a process modeled through entanglement percolation
\cite{acin2007entanglement, malik2022concurrence}. Therefore, as currently conceived, large-scale quantum communication
relies on two crucial steps \cite{perseguers2013distribution}: entanglement distribution, 
followed by manipulation
through local operations and classical communication (LOCC) \cite{chitambar2014everything} to 
establish end-to-end entanglement. 

Previous research on LOCC protocols has focused primarily on centralized and sequential approaches, where a designated entity coordinates entanglement manipulation and processes entangled pairs 
sequentially \cite{skrzypczykarchitecture, abane2025entanglement}. Parallel and distributed schemes, though still less explored, are increasingly recognized as 
promising alternatives that can enhance scalability and reduce communication overhead \cite{pirker2019quantum, pant2019routing}. In a realistic large-scale implementation of a decentralized protocol, 
competing demands among parties for entanglement raise fundamental questions about how entanglement resources can be efficiently allocated and utilized.

In this letter, we consider a decentralized scenario where multiple user pairs simultaneously request and establish end-to-end entanglement for private communication. 
The problem is formulated as a non-cooperative LOCC protocol in which user pairs independently seek to optimize their own end-to-end entanglement fidelity. The competition for limited entanglement resources gives rise to an emergent resource utilization mechanism shaped by the strategic choices user pairs make. We approach the problem from a game-theoretic perspective and focus on the resulting Nash equilibria (NEs),
where no user pair can increase its payoff by unilaterally changing its strategy.
Our results show that: (i) in quantum networks in which part or all of the initial entangled states are mixed, removing certain entanglement resources can increase the average end-to-end fidelity when users act non-cooperatively; (ii) the effect requires mixed states, but it peaks in networks containing both mixed and pure states, with the removed resources
that generate the strongest effect most often being Bell states;
and (iii) this effect is more pronounced in larger networks and when user pairs are distributed across multiple end nodes.

The effect described here can be interpreted as a quantum analog of selfish routing observed in classical communication and transportation networks, where agents choose paths that minimize their individual latency rather than the global performance \cite{wardrop1952road, koutsoupias1999worst}.
The inefficiency from such noncooperative behavior has been extensively studied as the origin of Braess's paradox in traffic networks, in which the addition of a road increases rather than decreases average travel time \cite{braess1968uber, braess2005on}. This paradox has been previously discussed in a four-node quantum network, where increasing the initial entanglement was shown to reduce the end-to-end concurrence \cite{banerjee2021braess}. To the best of our knowledge, the implications of noncooperative behavior remain unexplored in large, complex quantum communication networks. Here, we focus on complex quantum networks,
leading to results with important consequences for the efficient allocation of entanglement resources.

\begin{figure}[t]
    \centering
    \includegraphics[width=\columnwidth]{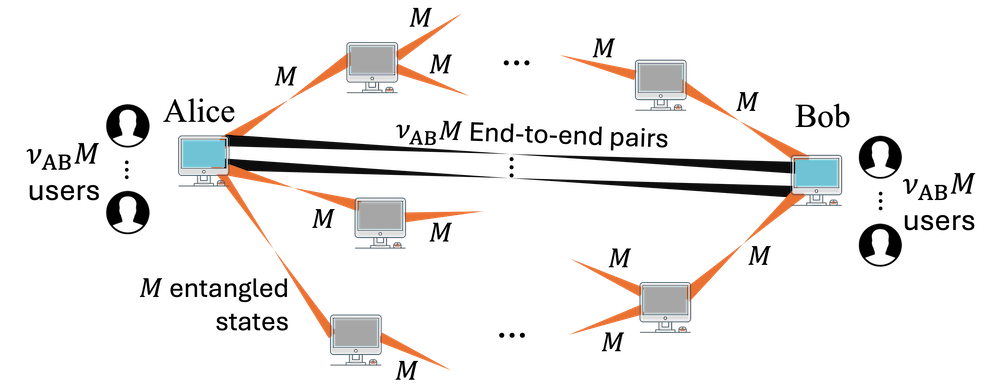}
    \caption{Schematic of a quantum communication network. Entanglement resources enable multiple users at the Alice node to establish independent end-to-end entangled states, along possibly different paths, with corresponding users at the Bob node. Orange links represent the initial entangled pairs, while the black links indicate the resulting Alice–Bob entangled states.
    }
    \label{schematic}
\end{figure}

\textit{Network setup and operations} -- We consider an arbitrary connected network consisting of 
$N$ nodes and $K$ edges. Nodes represent quantum devices capable of local qubit manipulation, and edges represent entanglement shared between neighboring nodes.
Initially, each edge is provided with $M$ identical copies of an entangled state. For concreteness, the entangled state on each edge is taken to be either the Bell state $|\Phi^+\rangle = \frac{1}{\sqrt{2}}\big(|00\rangle + |11\rangle\big)$, or a Werner state in the form
\begin{equation}
    \rho_{\text{W}}(p_0) = p_0|\Phi^+\rangle\langle\Phi^+|+\frac{1-p_0}{4}\mathds{1}_4.
\end{equation} 
The Werner parameter $p_0$ is linearly related to the fidelity, $F_0=\langle \Phi^+|\rho_{\text{W}}|\Phi^+\rangle=(3p_0+1)/4$.

We assume that the network has a specified pair of end nodes, Alice and Bob, each hosting  $\nu_{\text{AB}} M$ users that are paired one-to-one across the two ends (Fig.~\ref{schematic}). 
Here, $\nu_{\text{AB}}$ denotes the smaller of the degrees of the Alice and Bob nodes, where the degree of a node is the number of edges connected to it. This ensures that the demand scales with local connectivity, and thus the edges of the end nodes are not over- or under-utilized.
We also assume that individual user pairs establish end-to-end
entanglement by selecting a path of consecutive entangled states between Alice and Bob
that maximizes their own end-to-end fidelity. 
We denote by $\mathcal{R}_{\text{AB}}=\{\text{r}_1, \text{r}_2, ..., \text{r}_n\}$ the set of all Alice-Bob paths, where individual paths are required to be simple (i.e., no nodes or edges are repeated).
The operations described next are performed to establish the end-to-end entangled states while ensuring
non-preferential treatment across user pairs. 

If the demand on each edge matches the number of initial entangled states, 
\textit{entanglement swapping}  is performed at all intermediate nodes along each path \cite{zukowski1993bell}, consuming consecutive entangled states to establish direct end-to-end entanglement \cite{meng2025path}.
For an arbitrary path $r$, the Werner parameter of the resulting end-to-end entanglement is given by the product of the parameters along the path:
$p = \prod_{l\in r} p_l$. 
Below we use that this formula also applies to arbitrary pure states $|\phi (\lambda_0)\rangle = \sqrt{\lambda_0}|00\rangle + \sqrt{1-\lambda_0}|11\rangle$, which can be transformed into Werner states while preserving the fidelity $F_{\phi} = \sqrt{\lambda_0(1-\lambda_0)}+1/2$ \cite{bennett1996purification}, with the corresponding Werner parameter $p_{\phi} = (4F_{\phi}-1)/3=\big(4\sqrt{\lambda_0(1-\lambda_0)}+1\big)/3$. The efficiency of the entanglement swapping operation is assumed to be $1$. 

When there is a mismatch between the number of requested entangled pairs and the number of prepared states at a given edge, entanglement dilution or purification is used.
If $m>M$ entanglements are needed on an edge while only $M$ copies of Bell states are available, 
deterministic \textit{entanglement dilution} can convert the $M$ Bell states $|\Phi^+\rangle$ into $m$ copies of a pure state $|\phi (\lambda_0)\rangle$ in the asymptotic limit  $M\rightarrow \infty$ \cite{nielsen1999conditions}. The parameter $\lambda_0$ can be obtained from the conservation of the von Neumann entropy $S$, so that $S(\rho_{\phi})=M/m=-\lambda_0\text{log}_2\lambda_0 - (1-\lambda_0)\text{log}_2(1-\lambda_0)$.
On the other hand, no dilution protocol is available to address a deficit of Werner states.
Thus, if $m>M$ on a Werner-state edge, we simply use $p_0M/m$ as the expectation value of the Werner parameters, which uniquely defines the expectation value of the fidelity. 

In cases where $m<M$ on a Werner-state edge, 
the surplus of entangled pairs can be used to increase the fidelity of the remaining pairs via \textit{entanglement purification}. 
Specifically, we employ the non-deterministic BBPSSW protocol \cite{bennett1996purification}, which applies to the full range of fidelities. The protocol is implemented using a recurrence method, where 2-to-1 purification is iteratively applied to lowest-fidelity states until $m$ or $m + 1$ entangled states remain. In this process, the Werner states $\rho_1=\rho_{\text{W}}(p_1)$ and $\rho_2=\rho_{\text{W}}(p_2)$ are purified into $\rho=\rho_{\text{W}}(p)$, where 
\begin{equation}
    p = \frac{p_1+p_2+4p_1p_2}{3+3p_1p_2}, 
\end{equation}
and purification is successful with probability $(1+p_1p_2)/2<1$. 
Naturally, 
in cases where there is a surplus of Bell states, 
no further purification (or fidelity increase) is possible.


\begin{figure*}[t]
    \centering
    \includegraphics[width=1.95\columnwidth]{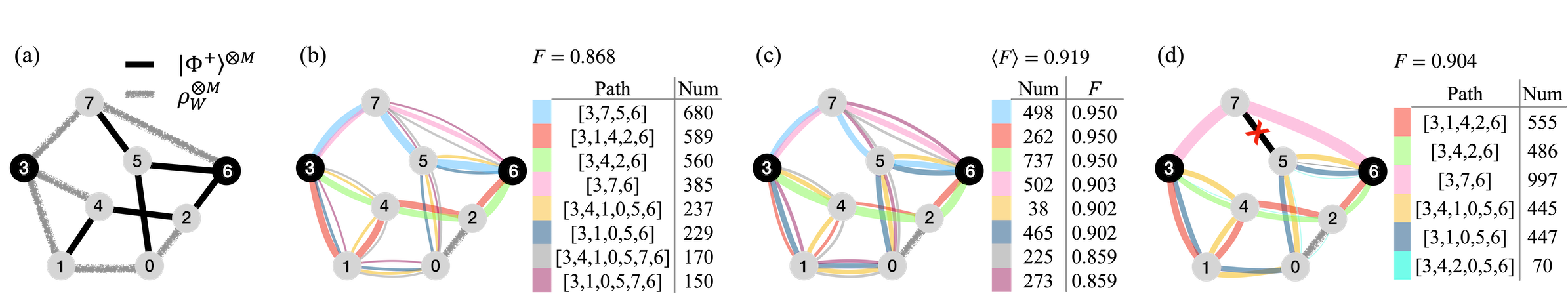}
    \caption{Noncooperative behavior in an example quantum network. (a) Network with Bell-state (black) and Werner-state (gray) edges in equal proportion, in which 3 and 6 serve as end nodes. (b–d) User pairs’ paths for establishing end-to-end entanglement at the NE (b), at the global optimum (c), and at the NE after removing edge (5,7) (d). The color-labeled tables list 
     the number of user pairs and/or fidelity associated with each path, 
    where the paths are the same for (b) and (c).
    The parameters are $N=8$,  $\nu_{\text{AB}}=3$, $M=1000$, and $F_0=0.95$.}
    \label{8-node example}
\end{figure*}
 
A state in which all user pairs independently maximize their end-to-end fidelity is an NE, where no user pair can further increase their fidelity by selecting a different path. 
In prospective applications of a quantum internet, each node represents a large number of users, which motivates us to focus on finite but large $M$ \cite{skrzypczykarchitecture}. 
Unless noted otherwise: 
i) we use $M=1000$; 
ii) the edge set is equally divided between Bell and Werner states, which, as we show, nearly maximizes the effect.

\textit{Comprehensive network example} -- 
We first illustrate the results using the 8-node network of uniform degree $3$ and Werner-state fidelity $F_0=0.95$ shown in Fig.~\ref{8-node example}(a).  
For nodes 3 and 6 as Alice and Bob, according to the protocol above, the $\nu_{\text{AB}} M=3M$ users establish direct entanglement through 8 distinct paths at the NE in Fig.~\ref{8-node example}(b), all with the same end-to-end fidelity $F = 0.868$ (to three decimal places). No user pair can improve their fidelity by switching to a different path, and yet this fidelity is appreciably lower 
than the maximum possible. 
The global optimum, defined as the state that maximizes the average end-to-end fidelity over all user pairs, reaches an average $\langle F\rangle = 0.919$
[Fig.~\ref{8-node example}(c)].
However, at the global optimum, nearly 
$1/6$ of all user pairs have fidelities lower than the NE fidelity, highlighting a drawback of the global optimum in failing to ensure fairness.
Crucially, removing the Bell states on edge $(5,7)$ increases the NE fidelity to $F= 0.904$ for all users [Fig.~\ref{8-node example}(d)]. The removal 
eliminates three of the original paths and results in a redistribution of other paths across the network. 
This counterintuitive fidelity improvement upon 
removal of entanglement resources constitutes a quantum manifestation of Braess’s paradox.

{\textit{Equilibrium dynamics}} --
In the limit of $M\rightarrow\infty $, an NE is called a Wardrop equilibrium (WE) \cite{wardrop1952road, Correa2011wardrop}. In this limit, the end-to-end fidelity is strictly the same along all chosen paths and is necessarily greater than or equal to that of any other path \cite{beckmann1956studies, Roughgarden2005selfish}.
Using finite but large $M$ accommodates the assumption of the large-$M$ limit in the Bell-state dilution described above and allows us to:

\smallskip
\noindent
($i$) approximate the WE by the NEs, which are less demanding to compute and can be obtained via a greedy algorithm [Supplemental Material (SM) \cite{sm}, Sec.~S1, 
where the algorithm's time complexity is also discussed]; 

\smallskip
\noindent
($ii$) retain the special properties of the WE not shared by the NEs, including its guaranteed uniqueness
(SM \cite{sm}, Sec.~S2).

\smallskip
\noindent
Accordingly, while the NE is generally not unique, they all approach a common equilibrium (the WE) for large $M$ (and we can informally treat them as singular).

\smallskip

We now present a framework for obtaining the WE through direct calculation, which is feasible for small networks and provides guidance for choosing $M$ in the NE approximation. Let $x_j$ denote the ratio between the number of user pairs choosing path $\text{r}_j$ and $M$, so that the set $\{x_j\}$ satisfies the normalization condition $\sum_{j=1}^n x_j = \nu_{\text{AB}}$. 
The normalized number of entangled states required on edge $e_i$ is then given by $y_i(x_1, ..., x_n) = \sum_{j, e_i \in \text{r}_j} x_j$, and the Werner parameter of the end-to-end entanglement generated along path $\text{r}_j$ is $p_{\text{r}_j} = \prod_{i, e_i \in \text{r}_j}g_{\alpha_i}(y_i)$,
where $\alpha_i \in \{1,2\}$ indicates whether edge $e_i$ is prepared in Bell states ($\alpha_i=1$) or mixed states ($\alpha_i=2$), and $g_1$ and $g_2$ are continuous functions that return the expected Werner parameter for 
each case. 
These functions are derived for the protocol above using the fact that dilution is done for Bell states when $y>1$ and purification is done for Werner states when $0<y<1$. 
The expressions for $g_{\alpha}(y)$ are derived in SM \cite{sm} (Sec.~S2).

Following the WE property that the end-to-end fidelity is the same for all used paths and not lower than for unused ones, we can construct the objective function
\begin{equation}
    \mathcal{C} (x_1, \dots, x_n) = \sum_{j=1}^n \left(\bar{p}-p_{\text{r}_j}\right)^2 x_j^{\beta_j}. 
    \label{eq:WE}
\end{equation}
Here, $\bar{p}$ is the (algebraic) average of the Werner parameters of the end-to-end entanglement of all user pairs,
\begin{equation}
    \bar{p} (x_1, \dots, x_n) = \frac{1}{\nu_{\text{AB}}}\sum_{j=1}^{n}p_{\text{r}_j} x_j,
    \label{eq:optimum}
\end{equation}
$\beta_j =
0 \, \text{if} \,\, p(\text{r}_j)\geq \bar{p}$, and
$\beta_j = 
1 \, \text{if} \,\, p(\text{r}_j)< \bar{p}$. The WE is given by 
solutions minimizing function $\mathcal{C}$ under the constraint $\sum_{j=1}^{n}x_j = \nu_{\text{AB}}$, where $\lim_{M\rightarrow\infty}\mathcal{C}=0$.

To quantify the extent of suboptimality emerging in the noncooperative protocol, we also calculate the global optimum state
by minimizing the objective function $\mathcal{C}_{\text{G}} (x_1, \dots, x_n)=-\bar{p}$.
In the results shown below, the optimizations of both $\mathcal{C}$ and $\mathcal{C}_{\text{G}}$ are computed using the basin-hopping method \cite{wales1997global}. The robustness of 
this
method is confirmed by benchmarking it against a convex method, as detailed in SM \cite{sm} (Sec.~S3). 
Furthermore,
NE simulations with $M=1000$ closely match the WE (SM \cite{sm}, Sec.~S4). 

\begin{figure}[t]
    \centering
    \includegraphics[width=0.9\columnwidth]{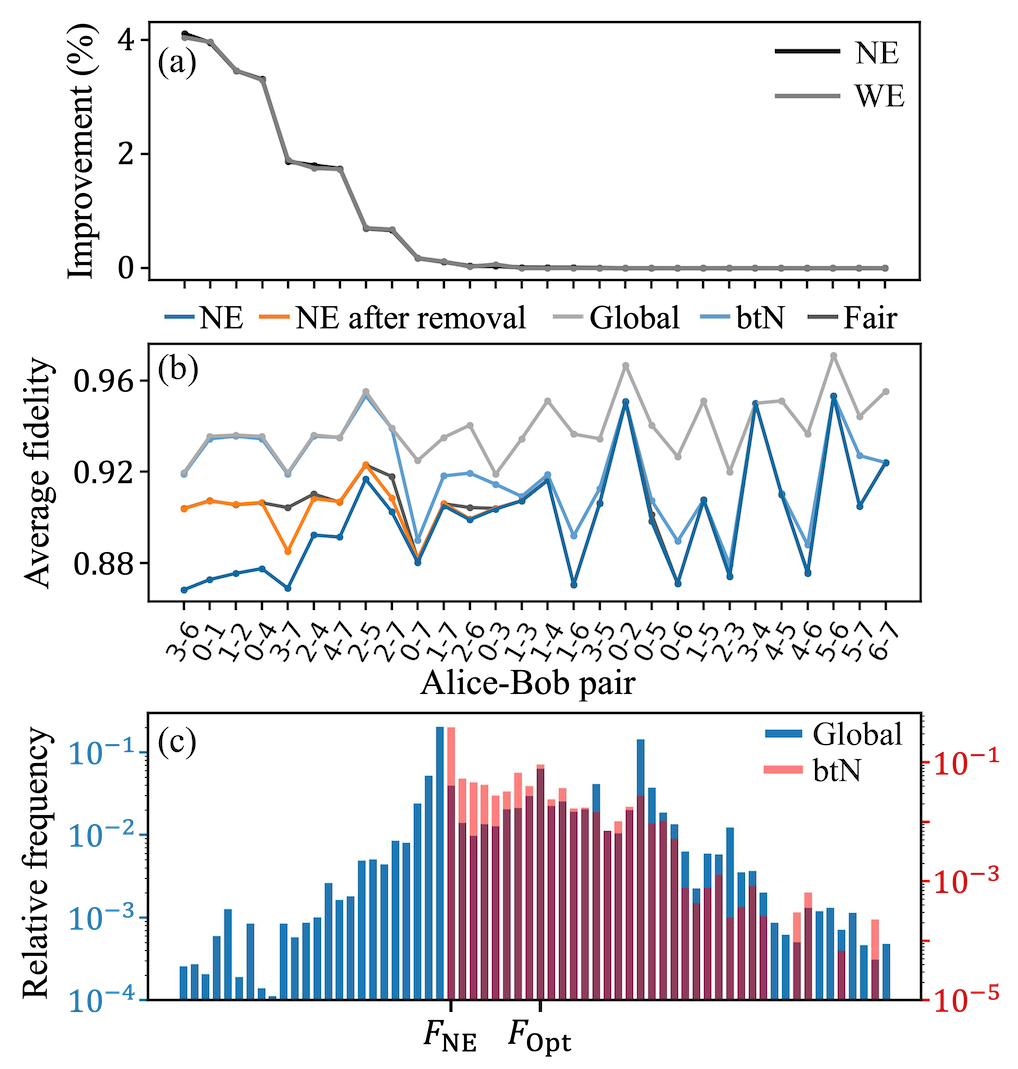}
    \caption{Equilibria and optima for all Alice-Bob pairs in the network of Fig.~\ref{8-node example}(a). (a) Largest fidelity improvement from single-edge removals for both NE and WE results. (b) Average end-to-end fidelity at the NE before and after edge removal, global optimum, btN optimum, and fair optimum. 
    (c) Histogram of the fidelity over all user pairs at the global and btN optima, scaled as $(F-F_{\text{NE}})/(F_{\text{Opt}}-F_{\text{NE}})$ for each Alice-Bob pair, where the indices indicate the NE value and global optimum average. 
    The parameters are the same as in Fig.~\ref{8-node example}. 
    } 
    \label{fig:example_all_AB}
\end{figure}

\textit{Contrasting equilibria and optima} -- 
We further explore the example network in Fig.~\ref{8-node example}(a) to illustrate the properties of the equilibrium states.
Figure~\ref{fig:example_all_AB}(a) shows the end-to-end fidelity improvement at the NE for all Alice-Bob pairs upon the one-edge removals with the most positive impact. Fidelity improvement is observed for more than 1/3 of the Alice-Bob pairs, and reaches a maximum of over $4\%$; for the others (not plotted), the impact is zero or slightly negative. The figure also shows excellent agreement between the WE results obtained through the basin-hopping method and the NE results from the greedy algorithm.
Details on the end-to-end fidelities of the NEs and global optimum are shown in Fig.~\ref{fig:example_all_AB}(b). 
For nearly all Alice-Bob pairs, the average fidelity at the global optimum is substantially higher than at the NE.
Figure~\ref{fig:example_all_AB}(c) shows, however, that the global optimum generally includes user pairs (36.5\% on average) with fidelity lower than at the NE, which confirms for other Alice-Bob pairs the result anticipated in  Fig.~\ref{8-node example}.
Moreover, even though part of the entangled states are not used at the global optimum [cf. Fig.~\ref{8-node example}(c)], merely removing unused entanglement does not generally bring the NE substantially closer to the global optimum (SM \cite{sm}, Sec.~S5).

It is important to examine the extent to which average fidelity and fairness can be improved simultaneously.
We therefore introduce two optimal solutions that seek to strike a compromise between the NE and the global optimum: 1) the {\it better-than-Nash} (btN) optimum, which maximizes the average end-to-end fidelity while requiring all user pairs to have fidelity not lower than the NE; 2) the {\it fair} optimum, which maximizes end-to-end fidelity under the constraint of being the same for all user pairs. See SM \cite{sm}, Sec.~S6, 
for details of the numerical implementation. 
As shown in Fig.~\ref{fig:example_all_AB}(b), both the btN optimum and fair optimum exhibit average end-to-end fidelity substantially higher than the NE for most Alice-Bob pairs. 
Figure~\ref{fig:example_all_AB}(c) further shows that the fidelity distribution for different user pairs in the btN optima is much narrower than the global optima on both sides of the distribution.
Note that the global optima and, under the given constraints, the btN optima are Pareto efficient, in a sense analogous to classical settings \cite{smith1979existence,cansever1986decentralized}, since no user pair can achieve higher fidelity by choosing a different path without reducing that of others.

\begin{figure*}
    \centering
    \includegraphics[width=2.05\columnwidth]{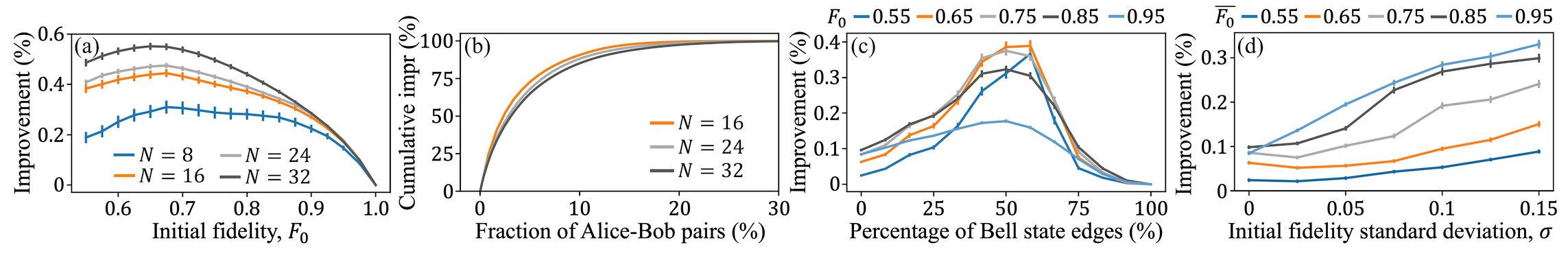}
\caption{End-to-end fidelity improvement in ER networks from optimal one-edge removals, averaged over all Alice-Bob pairs.  
(a) Effect as a function of the initial Werner-state fidelity for different network sizes. (b) Cumulative fidelity improvement over all Alice-Bob pairs ordered by effect size for the
16-, 24-, and 32-node networks in (a) with  $F_0=0.675$. 
(c) Effect as a function of the percentage of edges initially prepared in Bell states. (d) Effect in networks with only Werner states, 
where the initial fidelity $F_0$ of each edge is sampled from the discrete set $\{0.55,0.575, \ldots, 0.975\}$ with probabilities proportional to $e^{-[(F_0-\bar{F_0})/\sigma]^2/2}$ (a discrete truncated Gaussian). 
In (c-d), the network size is $N=16$. Each data point represents an average over 100 network realizations, 
where the bars indicate standard errors.
    }
    \label{fig:summary}
\end{figure*}

\textit{General networks} -- Through systematic analysis of the noncooperative protocol in random and regular quantum networks, we now show that the effect  
can be more pronounced in larger networks and for intermediate or low initial fidelity $F_0$.
Figure~\ref{fig:summary} summarizes the results for Erd\H os-R\'enyi (ER) random networks~\cite{erdos1960on}, generated by connecting each pair of nodes independently with probability $p=k/(N-1)$, where $k$ is the average degree. For concreteness, we consider networks of size $N=8$, $16$, $24$, and $32$ with $k=3$, under the constraint that the total number of edges is $K=Nk/2$. 

Figure~\ref{fig:summary}(a) shows how the extent of the effect consistently increases with network size. 
For each network size, it depends non-monotonically on $F_0$, peaking at $F_0\approx 0.669$ and vanishing for $F_0=1$ (when all edges are Bell states).  
The cumulative fidelity improvement upon edge removal over all possible pairs of Alice-Bob pairs shows that the percentage of pairs contributing to the effect increases with network size but is mainly concentrated in 20\% of all the pairs for the range of size considered [Fig.~\ref{fig:summary}(b)]. 
Similar results are also found in square lattice networks, as shown in SM \cite{sm} (Sec.~S7),
suggesting that they are not strongly dependent on the network structure.

The effect depends strongly on the proportion of Bell- versus Werner-state edges.
As shown in Fig.~\ref{fig:summary}(c) 
for 16-node ER networks, the effect peaks when the split is close to $50{:}50$, with the dependence of the peak on $F_0$ following a trend similar to the one in Fig.~\ref{fig:summary}(a).
While the effect vanishes 
when all edges are Bell states
(and the same would hold for non-maximally entangled pure states), 
it persists when all edges are Werner states and becomes more pronounced if their initial fidelities are allowed to be heterogeneous
[Fig.~\ref{fig:summary}(d)]; 
in this case, particularly when heterogeneity is high, the effect is enhanced for larger rather than smaller average $F_0$.
Moreover, in networks comprising both Bell and Werner states, the removed edges 
yielding greatest fidelity gains
are most often, but not always, Bell-state edges (SM \cite{sm}, Sec.~S8). 
This is because  such  
edges can
attract many paths by being locally less constraining while simultaneously increasing competition for entanglement resources along other segments of those paths.

\begin{figure}[b]
    \centering
    \includegraphics[width=0.99\columnwidth]{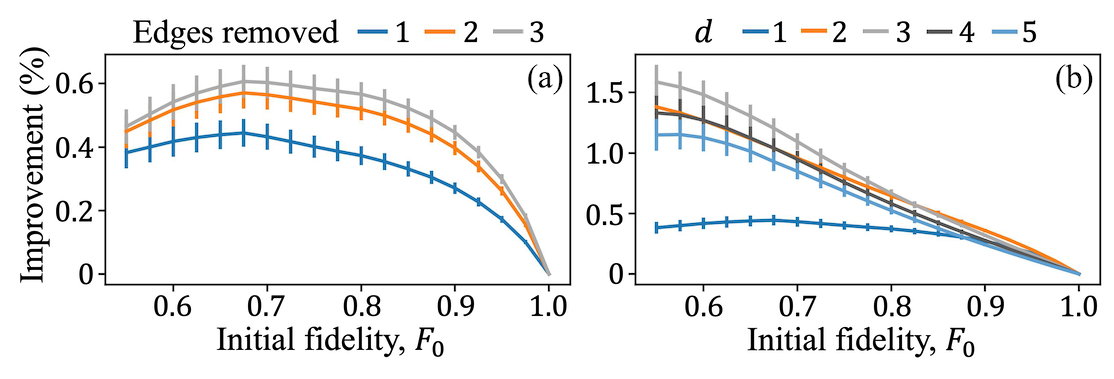}
    \caption{End-to-end fidelity improvement under optimal edge removals,
    for varying numbers of removals and Alice-Bob pairs.  
    (a) Improvement from optimal 2- or 3-edge removals compared with 1-edge removals [as in Fig.~\ref{fig:summary}(a)]. 
    (b) Improvement 
    from single-edge removals for networks with multiple Alice-Bob end nodes. }
    \label{fig:multiple_removal} 
\end{figure}

Thus far, we have focused on a single pair of Alice–Bob nodes, between which all user pairs attempt to establish entanglement, while considering the impact of single-edge removals. 
We now examine the persistence of the effect for multiple edge removals and multiple Alice–Bob pairs from which users can concurrently request end-to-end entanglement.
Figure~\ref{fig:multiple_removal}(a) shows that the average end-to-end fidelity can be further increased by removing more edges, thereby further reducing the initial entanglement budget [cf. Fig.~\ref{fig:summary}(a)].  
Extending to the case of $d$ Alice-Bob pairs, we assume $\nu_{\text{A}_i\text{B}_i} M/d$ user pairs between the $i$th Alice–Bob pair, where $\nu_{\text{A}_i\text{B}_i}$ denotes the smaller degree between nodes A$_i$ and B$_i$. 
For the representative case of $16$-node ER networks, the average fidelity improvement increases as $d$ rises from 1 to 3, before beginning to decrease for $d\ge 4$
[Fig.~\ref{fig:multiple_removal}(b)]. 
These results highlight the robustness of the effect under more general conditions.

\textit{Conclusion} -- 
Our analysis of noncooperative quantum communication protocols reveals that a quantum analog of selfish routing can lead to markedly suboptimal use of entanglement resources. Importantly, increasing the available entanglement may reduce the average end-to-end fidelity, while selective removal can improve it---a quantum manifestation of Braess’s paradox. 
Under the broad conditions considered here, the phenomenon arises from the presence of mixed entangled
states and vanishes for pure-only states. Surprisingly, though, the effect becomes most pronounced when mixed and pure states coexist in comparable numbers. Crucially, its impact also grows with network size and the number of distinct end nodes. 
Although noncooperative protocols are bound to converge to suboptimal equilibria, they retain the virtues of being decentralized and inherently egalitarian, since all users can achieve the same end-to-end fidelity.
Taken together, our results challenge the assumption that maximizing entanglement resources necessarily optimizes end-to-end fidelity, underscoring fundamental trade-offs in achieving fair and efficient entanglement utilization in quantum networks.

\let\oldaddcontentsline\addcontentsline
\renewcommand{\addcontentsline}[3]{}

\begin{acknowledgements}    

{\textit{Acknowledgements}} -- This work was supported by ARO MURI Grant No.~W911NF-21-1-0325. 

{\textit{Data availability} -- Code and data are available at \cite{github}. }

\end{acknowledgements}

\end{document}